\documentclass[english]{IEEEtran}
\usepackage{amsfonts}
\usepackage{amsgen,amsmath,amstext,amsbsy,amsopn,amssymb}
\usepackage{graphicx}
\usepackage{flushend}
\usepackage[usenames]{color}

\newtheorem{definition}{Definition}
\newtheorem{theorem}{Theorem}

\newtheorem{lemma}{Lemma}

\newcommand{\be}{\begin{equation}}
\newcommand{\ee}{\end{equation}}

\newcommand{\bbR}{\mathbb{R}}
\newcommand{\cB}{\mathcal{B}}

\newcommand{\RR}{\mathbb{R}}

\newcommand{\0}{{\mathbf{0}}}

\newcommand{\e}{{\mathbf{e}}}

\newcommand{\h}{{\mathbf{h}}}
\renewcommand{\u}{{\mathbf{u}}}
\renewcommand{\v}{{\mathbf{v}}}

\newcommand{\x}{{\mathbf{x}}}
\newcommand{\y}{{\mathbf{y}}}
\newcommand{\z}{{\mathbf{z}}}
\newcommand{\A}{{\mathbf{A}}}
\newcommand{\B}{{\mathbf{B}}}

\newcommand{\lLam}{{\mathbf{\Lambda}}}

\newcommand{\supp}{{\rm supp}}

\newcommand{\bbeta}{\boldsymbol{\beta}}

\newcommand{\diag}{{\rm diag}}

\allowdisplaybreaks

\begin{document}
\title{Sparse signal recovery by $\ell_q$ minimization under restricted isometry property
}
\author{Chao-Bing Song, Shu-Tao Xia}
\date{}
\maketitle
\begin{abstract}
In the context of compressed sensing, the nonconvex  $\ell_q$ minimization
with $0<q<1$ has been studied in recent years. In this paper, by generalizing the sharp bound for $\ell_1$ minimization of Cai and Zhang, we show that the condition $\delta_{(s^q+1)k}<\dfrac{1}{\sqrt{s^{q-2}+1}}$ in terms of \emph{restricted isometry constant (RIC)} can guarantee the exact recovery of $k$-sparse signals in noiseless case and the stable recovery of approximately $k$-sparse signals in noisy case by $\ell_q$ minimization.
This result is more general than the  sharp bound for $\ell_1$ minimization when the order of RIC is greater than $2k$ and illustrates the fact that a better approximation  to $\ell_0$ minimization is provided by $\ell_q$ minimization than that provided by $\ell_1$ minimization.
\end{abstract}
\renewcommand{\thefootnote}{\fnsymbol{footnote}} \footnotetext[0]{
This research is supported in part by the Major State Basic Research
Development Program of China (973 Program, 2012CB315803), the National
Natural Science Foundation of China (61371078), and the Research Fund for
the Doctoral Program of Higher Education of China (20130002110051).

All the authors are with the Graduate School at ShenZhen, Tsinghua University, Shenzhen, Guangdong 518055, P.R. China (e-mail: scb12@mails.tsinghua.edu.cn,  xiast@sz.tsinghua.edu.cn).} \renewcommand{\thefootnote}{\arabic{footnote}} \setcounter{footnote}{0}
\section{\label{sec:introduction}Introduction}
As a new paradigm for signal sampling, compressed sensing (CS) \cite{donoho2006compressed,candes2005decoding,candes2006robust}
has attracted a lot of attention in recent years. Consider a $k$-sparse signal
$\x=(x_1,x_2,\ldots,x_p)\in \bbR^p$ which has at most $k$ nonzero entries. Let
$\A\in\bbR^{n\times p}$ be a measurement matrix with $n\ll p$ and $\y=\A\x$ be
a measurement vector. CS deals with recovering the original signal $\x$ from
the measurement vector $\y$ by finding the sparsest solution to the underdetermined
linear system $\y=\A\x$, i.e., solving the following \emph{$\ell_0$ minimization}
problem:
\be
\min \|\x\|_0\qquad s.t.\qquad \A\x=\y, \label{eq:l_0}
\ee
where $\|\x\|_0:=|\{i:x_i\neq 0\}|$ denotes the $\ell_0$-norm of $\x$. Unfortunately,
as a typical combinatorial optimization problem, this optimal recovery algorithm
is NP-hard \cite{candes2005decoding}. One popular strategy is to relax the $\ell_0$ minimization
problem to an \emph{$\ell_1$ minimization} problem:
\be
\min \|\x\|_1\qquad s.t.\qquad \A\x=\y. \label{eq:l_1}
\ee
Due to the convex essence of $\ell_1$ minimization, we can solve it in polynomial time \cite{candes2005decoding}.

In order to describe the
equivalence condition between reconstruction algorithms with polynomial
time  and $\ell_0$ minimization, \emph{restricted isometry property (RIP)} is introduced in Cand{\`e}s and Tao \cite{candes2005decoding}, which has been one of the most popular properties of measurement matrix in CS.
We can rewrite the definition of RIP as follows.
\begin{definition}
\label{def:rip}
The measurement matrix $\A\in\mathbb{R}^{n\times p}$ is said to satisfy the $k$-order RIP if for any $k$-sparse signal $\mathbf{x}\in\mathbb{R}^{p}$,
\begin{equation}
(1-\delta)\Vert\mathbf{x}\Vert_2^2\le\Vert\A\mathbf{x}\Vert_2^2\le(1+\delta)\Vert\mathbf{x}\Vert_2^2, \label{eq:origin_def}
\end{equation}
where $0\le\delta\le1$. The infimum of $\delta$, denoted by $\delta_k$, is called the $k$-order \emph{restricted isometry constant (RIC)}  of $\A$.
When $k$ is not an integer, we define $\delta_k$ as $\delta_{\lceil k\rceil}$, where $\lceil\cdot\rceil$ denotes the ceiling function.
\end{definition}


There are a lot of papers to discuss the equivalence condition between $\ell_1$ minimization and $\ell_0$ minimization in terms of RIC, such as $\delta_k+\delta_{2k}+\delta_{3k}<1$ in Cand{\`e}s and Tao \cite{candes2005decoding}, $\delta_{2k}<\sqrt{2}-1$ in Cand{\`e}s \cite{candes2008restricted}, $\delta_{2k}<0.4652$ in Foucart \cite{foucart2010note}, $\delta_k<1/3$ in Cai and Zhang \cite{cai2012sharp}, and $\delta_{tk}<\sqrt{\frac{t-1}{t}}(t>4/3)$ in Cai and Zhang \cite{cai2013sparse}. In these conditions, $\delta_{k}+\delta_{2k}+\delta_{3k}<1$ is the first RIC condition, while $\delta_k<1/3$ and $\delta_{tk}<\sqrt{\frac{t-1}{t}}(t>4/3)$ are sharp bounds in the sense that we can find counterexample that $\ell_1$ minimization can't find $\x$ exactly if these conditions don't hold \cite{cai2012sharp}, \cite{cai2013sparse}.

Instead of $\ell_1$ minimization, from the fact that $\lim_{q\rightarrow 0}\|\x\|_q^q=\|\x\|_0$, solving an \emph{$\ell_q$($0<q<1$) minimization  } problem
\be
\min \|\x\|_q^q\qquad s.t.\qquad \A\x=\y \label{eq:l_q}
\ee
 may provide a better approximation to $\ell_0$ minimization.  The advantages of $\ell_q$
minimization can be found in \cite{lai2011unconstrained}. Although
finding a global minimizer of \eqref{eq:l_q} is NP-hard,  a lot of algorithms with polynomial time have been proposed to find
a local minimizer of \eqref{eq:l_q}, such as the algorithms in \cite{lai2011unconstrained}, \cite{daubechies2010iteratively}, \cite{chartrand2008iteratively}.

In practical applications, there often exist noises in measurements and the original signal $\x$ may be not exact sparse. In noisy case, we can relax the constraint in \eqref{eq:l_q} as follows,
\be
\min \|\x\|_q^q\qquad s.t.\qquad \y-\A\x\in\cB , \label{eq:l_q2}
\ee
where $\cB$ denotes some noise structure. In this setting, we need  to recover $\x$ with bounded errors, i.e., recover $\x$ stably.

Several RIC bounds of $\ell_q$ minimization are given in the literature, such as $\delta_{2k}<0.4531$ in Foucart and Lai \cite{foucart2009sparsest}, $\delta_{2k}<0.4931$ in Hsia and Sheu \cite{hsiaric}. Other similar results can be found in Saab, Chartrand and Yilmaz \cite{saab2008stable}, Lai and Liu \cite{lai2011new}, Zhou Kong, Luo and Xiu \cite{zhou2013new}.
In this paper, we mainly focus on the RIC condition of $\ell_q$ minimization.
We show that if $\delta_{(s^q+1)k}<\frac{1}{\sqrt{s^{q-2}+1}}(s>0)$,  $\ell_q$ minimization can recover $k$-sparse signal exactly in noiseless case and recover approximately $k$-sparse signal stably in noisy case. From this condition, we show that as a relaxtion way closer to $\ell_0$ minimization, $\ell_q$ minimization can guarantee sparse signal recovery in a more general condition in terms of RIC.

The remainder of the paper is organized as follows. In Section \ref{sec:preliminaries}, we introduce
related notations and lemmas. In Section \ref{sec:main_results}, we give our main results  in both noiseless and noisy settings. In Section \ref{sec:proof}, unified proofs are given to the main results in Section \ref{sec:main_results}. Finally,  conclusion is given
in Section \ref{sec:conclusion}.
\section{\label{sec:preliminaries}Preliminaries}
Let $\e_i$'s $\in\bbR^p$ are different unit vectors with one entry of $1$ or $-1$ in position $i\in\{1,2,\ldots,p\}$ and other entries
of zeros, which Cai and Zhang \cite{cai2012sharp}  call indicator vectors. Let $\v=\sum_{i=1}^{p}v_i\e_i$ be an arbitrary vector in $\bbR^p$, where $\forall i\in\{1,2,\ldots,p\}, v_i\ge0$.
Let $\supp(\v)$ denote the support of $\v$ or the set of indices of nonzero entries in $\v$.
Let $\v_{\max(k)}$  be the vector $\v$ with all but the largest $k$ entries in absolute values set to zeros
and $\v_{-\max(k)}=\v-\v_{\max(k)}$. For $0<q<\infty$, let $\ell_q$-norm of a vector $\v\in\bbR^p$
as $\|\v\|_q=(\sum_{i=1}^{p}|v_i|^q)^{1/q}$. In addition, let $\|\v\|_{\infty}=\sup_i |v_i|$ and
$\|\v\|_0=|\supp(\v)|$ be the number of nonzero entries in $\v$. Let $\v^q=\sum_{i=1}^{p}v_i^q\e_i$
be ``\emph{the $q$ power of the vector} $\v$''. In addition, let
$\sigma(\A)$ denote the spectral norm of $\A$.

Then we introduce direct consequences of the H$\ddot{o}$lder inequality as follows.
\begin{lemma}
\label{lem:holder}
If $\forall \v\in\bbR^p$ and $0<q<1$, \[
\|\v\|_q\le p^{\frac{1}{q}-\frac{1}{2}}\|\v\|_2.\]
Moreover, if $\v$ is $k$-sparse, then \[
\|\v\|_q\le k^{\frac{1}{q}-\frac{1}{2}}\|\v\|_2.\]
\end{lemma}

The following lemma introduced in Cai and Zhang \cite{cai2013sparse} is crucial to get the proposal results on $\delta_{(s^q+1)k}$ .
\begin{lemma}[Sparse Representation of a Polytope]
\label{lm:mean}
For a positive number $\alpha$ and a positive integer $t$, define the polytope $T(\alpha, t) \subset \RR^p$ by
\[
T(\alpha, t) =\{\v\in \mathbb{R}^p: \|\v\|_\infty\leq \alpha, \; \|\v\|_1\leq t\alpha\}.
\]
For any $\v\in \mathbb{R}^p$, define the set of sparse vectors $U(\alpha, t, \v) \subset \RR^p$ by
\begin{eqnarray}
\label{eq:mean_u}
U(\alpha, t, \v) =\{\u\in \RR^p:\;  {\rm supp}(\u) \subseteq {\rm supp}(\v),\;  \|\u\|_0\le t, \nonumber\\
 \|\u\|_1=\|\v\|_1,\; \|\u\|_\infty\leq \alpha\}.
\end{eqnarray}
Then $\v \in T(\alpha, t)$ if and only if $\v$ is in the convex hull of  $U(\alpha, t, \v)$. In particular, any $\v\in T(\alpha, t)$ can be expressed as
\begin{eqnarray}
\v=\sum_{i=1}^N\lambda_i \u_i, \quad\mbox{and }\; 0\leq \lambda_i\leq 1,\quad \sum_{i=1}^N \lambda_i=1,\nonumber\\
\quad\mbox{and } \; \u_i \in U(\alpha, t, \v).
\end{eqnarray}
\end{lemma}
\section{\label{sec:main_results} Main Results}
In noiseless case, we have the following result.
\begin{theorem}
\label{thm:1} Assume that $\x\in\bbR^{p}$ is $k$-sparse signal and $\y=\A\x$ with $\y\in\bbR^{n}, \A\in\bbR^{n\times p}$. Then if the $(s^q+1)k$-order RIC of the measurement matrix $\A$ satisfies
\be
\delta_{(s^q+1)k}<\frac{1}{\sqrt{s^{q-2}+1}}, \label{eq:ric}
\ee
the minimizer $\hat{\x}$ of \eqref{eq:l_q}  will recover $\x$ exactly.
\end{theorem}

In noisy case, two types of bounded noisy setting
\begin{itemize}
 \item $\cB=\{\z:\|\z\|_2\le\eta\}$,
\item $\cB=\{\z:\|\A^{\! T}\z\|_{\infty}\le\eta\}$,
\end{itemize}
are of particular interest. The first bounded noise setting was introduced in \cite{donoho2006stable}. The second one was motivated by Dantzig Selector in \cite{candes2007dantzig}. 
The corresponding results in the two noisy cases are given in Theorems \ref{thm:2} and \ref{thm:3}, respectively.
\begin{theorem}
\label{thm:2} Assume that $\x\in\bbR^{p}$ is approximately $k$-sparse signal, $\y=\A\x+\z$ with $\y,\z\in\bbR^{n}, \A\in\bbR^{n\times p},\|\z\|_2\leq\epsilon$, and $\cB=\{\z:\|\z\|_2\le\eta\}$ with $\eta\ge\epsilon+\sigma(\A)\|\x_{\overline{T}}\|_2$ in \eqref{eq:l_q2}. Then if the $(s^q+1)k$-order RIC of the measurement matrix $\A$ satisfies
\be
\delta_{(s^q+1)k}<\frac{1}{\sqrt{s^{q-2}+1}},\nonumber
\ee
the minimizer $\hat{\x}$ of \eqref{eq:l_q2} will recover x stably as follows:
\begin{eqnarray}
\|\hat{\x}-\x\|_2&\leq&\frac{\sqrt{2(1+\delta_{(s^q+1)k})}\;(\epsilon+\eta)}{1-\sqrt{s^{q-2}+1}\;\delta_{(s^q+1)k}} \nonumber\\
&&\!\!\!\!\!\!\!\!\!\!\!\!\!\!\!\!\!\!\!\!\!+\left(\frac{\sqrt{2(1+\delta_{(s^q+1)k})}\;\sigma(\A)}{1-\sqrt{s^{q-2}+1}\;\delta_{(s^q+1)k}}+1\right)\|\x_{-\max(k)}\|_2.\label{eq:th2}
\end{eqnarray}
\end{theorem}
\begin{theorem}
\label{thm:3} Assume that $\x\in\bbR^{p}$ is approximately $k$-sparse signal, $\y=\A\x+\z$ with $\y,\z\in\bbR^{n}, \A\in\bbR^{n\times p},\|\A^{\!T}\z\|_{\infty}\leq\epsilon$, and $\cB=\{\z:\|\A^{\!T}\z\|_{\infty}\le\eta\}$ with $\eta\ge\epsilon+\sigma^2(\A)\|\x_{\overline{T}}\|_2$  in \eqref{eq:l_q2}. Then if the $(s^q+1)k$-order RIC of the measurement matrix $\A$ satisfies
\be
\delta_{(s^q+1)k}<\frac{1}{\sqrt{s^{q-2}+1}}, \nonumber
\ee
the minimizer $\hat{\x}$ of \eqref{eq:l_q2} will recover x stably as follows:
\begin{eqnarray}
\|\hat{\x}-\x\|_2&\leq&\frac{\sqrt{2(s^q+1)k}\;(\epsilon+\eta)}{1-\sqrt{s^{q-2}+1}\;\delta_{(s^q+1)k}} \nonumber\\
&&\!\!\!\!\!\!\!\!\!\!\!\!\!\!\!\!\!\!\!\!\!\!\!\!\!+\left(\frac{\sqrt{2(s^q+1)k}\;\sigma^2(\A)}{1-\sqrt{s^{q-2}+1}\;\delta_{(s^q+1)k}}+1\right)\|\x_{-\max(k)}\|_2.\label{eq:th3}
\end{eqnarray}
\end{theorem}

The proposed RIC condition is a natural generalization of the sharp result $\delta_{tk}<\sqrt{\frac{t-1}{t}}=\frac{1}{\sqrt{(t-1)^{-1}+1}}(t>4/3)$ in Cai and Zhang \cite{cai2013sparse}. Rewrite $\delta_{tk}<\frac{1}{\sqrt{(t-1)^{1-\frac{2}{q}}+1}}$ for \eqref{eq:ric}, and it is easy to find that $\frac{1}{\sqrt{(t-1)^{1-\frac{2}{q}}+1}}<\frac{1}{\sqrt{(t-1)^{-1}+1}}$ if $0<q<1$ and $t>2$. Therefore, in terms of RIC with order more than $2k$,  the condition of the measurement matrix $\A$ is relaxed if we use $\ell_q(0<q<1)$ minimization instead of $\ell_1$ minimization. In addition, in Theorems \ref{thm:2} and \ref{thm:3}, we use a relatively stricter condition $\eta\ge\epsilon+\sigma(\A)\|\x_{\overline{T}}\|_2$ and $\eta\ge\epsilon+\sigma^2(\A)\|\x_{\overline{T}}\|_2$ respectively than $\eta\ge\epsilon$ used in Cai and Zhang \cite{cai2013sparse}. In our proofs, in order to get an analytic upper bound of $\|\hat{\x}-\x\|_2$, the stricter condition may be necessary.
Finally, although the proposed bound is better than the existing results, a further research is still needed to verify whether it is sharp or not.
%
\medskip\noindent

\section{\label{sec:proof}Proofs}

In this section, firstly, our proofs are stated in general case. Then three cases including a noiseless case and two noise cases are discussed separately.
\begin{proof}
Assume that $\x$ is approximately $k$-sparse signal. Let $T$ denote the support of the largest $k$ entries of $\x$ and $\overline{T}$ denote the complement of $T$ . Let $\x_T(\x_{\overline{T}})$ denote the vector that sets all entries of $\x$ but the entry in $T(\overline{T})$ to zero. Let $\e^{\prime}=\A\x_{\overline{T}}+\e$, and we have $\y=\A\x_{T}+\e^{\prime}$. Assume that $\y-\A\x_{T}\in\cB$ and $\hat{\x}$ is the minimizer of \eqref{eq:l_q2}. Let $\hat{\x}=\x_T+\h$, and we have
\be
\|\x_T\|_q^q-\|\h_T\|_q^q+\|\h_{\overline{T}}\|_q^q\le\|\x_T+\h\|_q^q\le\|\x_T\|_q^q.\nonumber
\ee
Immediately,
\be
\|\h_{-\max(k)}\|_q^q\le\|\h_{\overline{T}}\|_q^q\le\|\h_{T}\|_q^q\le\|\h_{\max(k)}\|_q^q.\label{eq:h}
\ee

Note that from the definitions in Section \ref{sec:preliminaries} and the beginning of the proof, $\x_{T}(\x_{\overline{T}})$ is equivalent to $\x_{\max(k)}(\x_{-\max(k)})$, introducing the symbol $T(\overline{T})$ is just for distinguishing $\h_{T}(\h_{\overline{T}})$ from $\h_{\max(k)}(\h_{-\max(k)})$.

 Then,  assume that $ks^{q}$ is an integer.
 Let $\h=\sum_{i=1}^{p}h_i\e_i$,
where $\e_i$'s  are indicator vectors. Without loss of generality, assume that
 $h_1\ge h_2\ge\cdots\ge h_p\ge0$.
Set $\alpha^q=\|\h_{\max(k)}^q\|_1\,/k$. We divide $\h_{-\max(k)}$ into two  parts with disjoint supports,
$\h_{-\max(k)}=\h_1+\h_2$, where $$\h_1=\h\cdot 1_{\{i:|\h_{-\max(k)}(i)|> \alpha/s\}},
\quad \h_2=\h\cdot 1_{\{i:|\h_{-\max(k)}(i)|\leq \alpha/s\}}.$$
Then $\h_{-\max(k)}^q=\h_1^q+\h_2^q$, $\|\h_1^q\|_1\leq \|\h_{-\max(k)}^q\|_1\leq  k \alpha^{q}$; besides, all non-zero entries of $\h_1^q$  has magnitude larger than $(\alpha/s)^q$, so $\h_1^q$ is $ks^{q}$-sparse. Let $|\text{supp}(\h_1^q)|=m$, then
\begin{equation}\label{eq:main1}
\begin{split}
\|\h_2^q\|_1& =\|\h_{-\max(k)}^q\|_1-\|\h_1^q\|_1  \leq k\alpha^q - \frac{m\alpha^q}{s^q}\\
&=(ks^q-m)\cdot (\frac{\alpha}{s})^q, \\
\|\h_2^q\|_\infty & \leq (\frac{\alpha}{s})^q.
\end{split}
\end{equation}

We now apply Lemma \ref{lm:mean}. Then $\h_2^{q}$  can be expressed as a convex combination of sparse vectors: $\h_2^q=\sum_{i=1}^N\lambda_i \u_i^q$, where $\u_i$ is $(ks^q-m)$-sparse.
Now we suppose $\mu\geq0, c\geq0$ are to be determined. Denote $\bbeta_i^q=\h_{\max(k)}^q+\h_1^q+\mu \u_i^q$, then
\begin{eqnarray}\label{eq:mainth2}
&&\sum_{j=1}^N\lambda_j \bbeta_j^q-c\bbeta_i^q\nonumber\\
&=&\h_{\max{(k)}}^q+\h_1^q+\mu \h_2^q-c\bbeta_i^q \nonumber\\
 &=& (1-\mu-c) (\h_{\max(k)}^q+\h_1^q)-c\mu \u_i^q+\mu \h^q.
\end{eqnarray}
and $\bbeta_i^q$, $\sum_{j=1}^N\lambda_j \u_j^q-c\bbeta_i^q-\mu \h^q$ are all $(s^q+1)k$-sparse vectors.

Define $\lLam:=\diag(h_1^{1-q},h_2^{1-q},\ldots,h_p^{1-q}),\B:=\A\lLam$. Then $\B\h^{q}=\A\lLam \h^{q}=\A\h=\mathbf{0}$.

We can check the following identity in $\ell_2$ norm,
\begin{equation}\label{eq:thmain_identity}
\begin{split}
&\sum_{i=1}^N\lambda_i\|\B(\sum_{j=1}^N\lambda_j\bbeta_j^q -c\bbeta_i^q)\|_2^2  \\
&\qquad+ (1-2c)\sum_{1\leq i<j\leq N}\lambda_i\lambda_j\|\B(\bbeta_i^q-\bbeta_j^q)\|_2^2 \\
&=\sum_{i=1}^N \lambda_i(1-c)^2\|\B\bbeta_i^q\|_2^2.
\end{split}
\end{equation}
Since $\B\h^q=\0$ and \eqref{eq:mainth2}, we have
\begin{eqnarray}
&&\B(\sum_{j=1}^N\lambda_j\bbeta_j^q - c\bbeta_i^q) \nonumber\\
&=& \B((1-\mu-c)(\h_{\max(k)}^q+\h_1^q)-c\mu \u_i^q+\mu \h^q)\nonumber\\
&=&\A((1-\mu-c)\lLam(\h_{\max(k)}^q+\h_1^q)-c\mu \lLam \u_i^q+ \mu \h)\nonumber\\
&=&\A((1-\mu-c)(\h_{\max(k)}+\h_1)-c\mu \lLam \u_i^q+\mu \h).\nonumber\\
&&\B\bbeta_i^q\nonumber\\
&=&\A(\lLam(\h_{\max(k)}^q+\h_1^q)+\mu\lLam \u_i^q)\nonumber\\
&=&\A(\h_{\max(k)}+\h_1+\mu\lLam \u_i^q).\nonumber
\end{eqnarray}

Assume that
 \be
 \langle\A(\h_{\max(k)}+\h_1),\A\h\rangle\le\rho\|\h_{\max(k)}+\h_1\|_2 \label{eq:proof2}
 \ee with some $\rho\ge0$.
 Set $c=\frac{1}{2}$, $\mu=\frac{-1+\sqrt{s^{q-2}+1}}{s^{q-2}}$. For notational convenience, we write $\delta$ for $\delta_{(s^q+1)k}$. Let the left-hand side of \eqref{eq:thmain_identity} minus the right-hand side, we get
 \begin{eqnarray}
 0&=&\sum_{i=1}^{N}\lambda_i\|\B(\sum_{j=1}^{N}\lambda_j\bbeta_j^q-c\bbeta_i^q)\|_2^2\nonumber\\
 &&-\sum_{i=1}^{N}\lambda_i(1-c)^2\|\B\bbeta_i^q\|_2^2 \nonumber\\
  &=&\!\!\!\!\!\sum_{i=1}^{N}\lambda_i\|\A((1-\mu-c)(\h_{\max(k)}+\h_1)-c\mu\Lambda\u_i^q+\mu\h)\|_2^2\nonumber\\
  &&-\sum_{i=1}^{N}\lambda_i(1-c)^2\|\A(\h_{\max(k)}+\h_1+\mu\Lambda\u_i^q)\|_2^2\nonumber\\
  &=&\sum_{i=1}^{N}[\|\A((1-\mu-c)(\h_{\max(k)}+\h_1)
  -c\mu\Lambda\u_i^q)\|_2^2\nonumber\\
  &&+2\langle\A((1-\mu-c)(\h_{\max(k)}+\h_1)-c\mu\Lambda\u_i^q),\mu\A\h\rangle\nonumber\\
  &&+\|\mu\A\h\|_2^2]-\sum_{i=1}^{N}\lambda_i(1-c)^2\nonumber\\
  &&\cdot(\|\A(\h_{\max(k)}+\h_1+\mu\Lambda\u_i^q)\|_2^2)\nonumber\\
  &\le&\sum_{i=1}^{N}\lambda_i[(1+\delta)((1-\mu-c)^2\|\h_{\max(k)}+\h_1\|_2^2\nonumber\\
  &&+c^2\mu^2\|\Lambda\u_i^q\|_2^2)]+\|\mu\A\h\|_2^2\nonumber\\
  &&+2\langle\A((1-\mu-c)(\h_{\max(k)}+\h_1-c\mu\Lambda\h_2^q),\mu\A\h\rangle\nonumber\\
  &&\!\!\!\!\!-\sum_{i=1}^{N}\lambda_i(1-\delta)(1-c)^2(\|\h_{\max(k)}+\h_1\|_2^2+\mu^2\|\Lambda\u_i^q\|_2^2)\nonumber\\
  &=&\!\!\!\!\!\sum_{1=1}^{N}\lambda_i[(1+\delta)((\frac{1}{2}-\mu)^2\|\h_{\max(k)}+\h_1\|_2^2+\frac{1}{4}\mu^2\|\Lambda\u_i^q\|_2^2]\nonumber\\
  &&+\langle\A((1-\mu)(\h_{\max(k)}+\h_1)),\mu\A\h\rangle\nonumber\\
  &&-\sum_{i=1}^{N}\frac{1}{4}\lambda_i(1-\delta)(\|\h_{\max(k)}+\h_1\|_2^2+\mu^2\|\Lambda\u_i^q\|_2^2)\nonumber\\
   &\leq&((\frac{1}{2}-\mu+(\frac{1}{2}s^{q-2}+1)\mu^2)\delta-\mu+\mu^2)\nonumber\\
  &&\cdot \|\h_{\max(k)}+\h_1\|_2^2+\mu(1-\mu)\rho\|\h_{\max(k)}+\h_1\|_2\label{eq:proof41}\nonumber\\
 &=&(\sqrt{s^{q-2}+1}(\mu-\mu^2)\delta-(\mu-\mu^2))\|\h_{\max(k)}+\h_1\|_2^2\nonumber\\
 &&+(\mu-\mu^2)\rho\|\h_{\max(k)}+\h_1\|_2.\!\!\!\!\!\!\!\!\!\!\label{eq:proof4}
 \end{eqnarray}

 Consider $\|\h_{\max(k)}+\h_1\|_2$ as the independent variable in the inequality \eqref{eq:proof4}$\ge 0$. If we want the solution about $\|\h_{\max(k)}+\h_1\|_2$ is upper bounded, the coefficient of the second-order term should be less than zero. Therefore, we have
 \be
 \delta<\frac{1}{\sqrt{s^{q-2}+1}},
 \ee
 and
 \be
 \|\h_{\max(k)}+\h_1\|_2\le\frac{\rho}{1-\sqrt{s^{q-2}+1}\;\delta}.\label{eq:proof5}
 \ee

 In \eqref{eq:proof41}, we used the fact that
\begin{eqnarray}
\quad \|\lLam \u_i^q\|_2^2&\leq& \sum_{j=k+m+1}^{(s^q+1)k} (|h_j|^{1-q}\,\|\u_{i}^q\|_{\infty})^2\nonumber\\
&\le& (ks^q-m)((\frac{\alpha}{s})^{1-q}(\frac{\alpha}{s})^{q})^2 \label{eq:1}\\
  &\le& ks^{q-2}\alpha^2 \nonumber\\
  &=&ks^{q-2}\left(\frac{\|\h_{\max(k)}^q\|_1^{1/q}}{k^{1/q}}\right)^2\nonumber\\
  &=&ks^{q-2}\left(\frac{\|\h_{\max(k)}\|_q}{k^{1/q}}\right)^2\nonumber\\
  &\le& ks^{q-2}\left(\frac{k^{1/q-1/2}\|\h_{\max(k)}\|_2}{k^{1/q}}\right)^2 \label{eq:2}\\
  &\le&
  s^{q-2}\|\h_{\max(k)}+\h_1\|_2^2, \nonumber                                                                                                                                                                                                                                                                                                                                                                                                                                                                                                                                                                                                                                                                                                                                                                                                                                                                                                                                                                                                                                                                                                                                                                                                                                                                                                                                                                                                                                                                                                                                                                                                                                                                                                                                                                                                                                                                                                                                                                                                                                                                                                                                     \end{eqnarray}
where \eqref{eq:1} is from \eqref{eq:main1} and \eqref{eq:2} is from Lemma \ref{lem:holder}.

If $(s^q+1)k$ is not an integer, note $(s^{\prime})^q = \lceil s^q k \rceil/k$, then $s^{\prime}>s$, $k(s^{\prime})^q$ is an integer, from the above derivations, we know that if
$$\delta=\delta_{(s^q+1)k}=\delta_{((s^{\prime})^q+1)k} < \frac{1}{\sqrt{(s^{\prime})^{q-2}+1}},$$
\eqref{eq:proof5} holds. While\[
\frac{1}{\sqrt{s^{q-2}+1}}<\frac{1}{\sqrt{(s^{\prime})^{q-2}+1}},\]
so if $(s^q+1)k$ is not an integer, the condition $\delta_{(s^q+1)k}<\frac{1}{\sqrt{s^{q-2}+1}}$ is still enough to guarantee that the solution about $\|\h_{\max(k)+\h_1}\|_2$ of the inequality \eqref{eq:proof4}$\ge0$ is upper-bounded.

 From \cite[Lemma 5.4]{cai2012sharp} and \eqref{eq:h}, we have $\|\h_{-\max(k)}\|_2\le\|\h_{\max(k)}\|_2$. So
 \begin{eqnarray}
 \|\hat{\x}-\x_{\max(k)}\|_2=\|\h\|_2&=&\sqrt{\|\h_{\max(k)}\|_2^2+\|\h_{-\max(k)}\|_2^2} \nonumber\\
 &\le&\sqrt{2}\|\h_{\max(k)}\|_2\nonumber\\
 &\le&\sqrt{2}\|\h_{\max(k)}+\h_1\|_2.\nonumber
 \end{eqnarray}
 Then
 \begin{eqnarray}
 \|\hat{\x}-\x\|_2&\le&\|\hat{\x}-\x_{\max(k)}\|_2+\|\x_{-\max(k)}\|_2\nonumber\\
&\le&\sqrt{2}\|\h_{\max(k)}+\h_1\|_2+\|\x_{-\max(k)}\|_2\nonumber\\
&\le&\frac{\sqrt{2}\rho}{1-\sqrt{s^{q-2}+1}\;\delta}+\|\x_{-\max(k)}\|_2. \label{eq:proof3}
 \end{eqnarray}

Next, we discuss the noiseless case and the two noisy cases respectively.

 \begin{enumerate}
 \item The noiseless case: If $\x$ is $k$-sparse, then $\A\h=\A\hat{\x}-\A\x_{\max(k)}=\A\hat{\x}-\A\x=\0$. Therefore in \eqref{eq:proof2}, let $\rho=0$, then in \eqref{eq:proof3}, we have $\|\hat{\x}-\x\|_2=0$, i.e., $\hat{\x}$ recovers $\x$ exactly. This completes the proof of Theorem \ref{thm:1}.
\item The noisy case $\cB=\{\z:\|\z\|_2\le\eta$\}: If $\x$ is approximately $k$-sparse, $\|\y-\A\x\|_2\le\epsilon$, and the spectral norm of $\A$ is $\sigma(\A)$, then
\begin{eqnarray}
&&\langle\A(\h_{\max(k)}+\h_1),\A\h\rangle\nonumber\\
&\le&\|\A(\h_{\max(k)}+\h_1)\|_2\|\A\h\|_2\nonumber\\
&\le&\sqrt{1+\delta}\|\h_{\max(k)}+\h_1\|_2(\|\y-\A\hat{\x}\|_2\nonumber\\
&&+\|\y-\A\x\|_2+\|\A\x_{-\max(k)}\|_2)\nonumber\\
&\le&\!\!\!\!\!\sqrt{1+\delta}(\eta+\epsilon+\sigma(\A)\|\x_{-\max(k)}\|_2)\nonumber\\
&&\cdot\|\h_{\max(k)}+\h_1\|_2.
\end{eqnarray}
In this case, the assumption $\|\y-\A\x_T\|_2\in\cB$ holds if $\eta\ge\epsilon+\sigma(\A)\|\x_{-\max(k)}\|_2.$
Therefore, in \eqref{eq:proof2}, let $\rho=\sqrt{1+\delta}(\epsilon+\eta+\sigma(\A)\|\x_{-\max(k)}\|_2)$, then  we have \eqref{eq:th2} from \eqref{eq:proof3}. This proves Theorem \ref{thm:2}.
\item The noisy case $\cB=\{\z:\|\A^{\! T}\z\|_{\infty}\le\eta\}$: If $\x$ is approximately $k$-sparse, $\|\A^{\! T}(\y-\A\x)\|_{\infty}\le\epsilon$, the spectral norm of $\A$ is $\sigma(\A)$, then
\begin{eqnarray}
&&\langle\A(\h_{\max(k)}+\h_1),\A\h\rangle\nonumber\\
&=&\langle\h_{\max(k)}+\h_1,\A^{\! T}\A\h\rangle\nonumber\\
&\le&\|\h_{\max(k)}+\h_1\|_1\cdot\|\A^{\! T}\A\h\|_{\infty}\nonumber\\
&=&\|\h_{\max(k)}+\h_1\|_1\cdot\|\A^{\! T}\A(\hat{\x}-\x_{\max(k)})\|_{\infty}\nonumber\\
&\le&\sqrt{(s^q+1)k}\|\h_{\max(k)}+\h_1\|_2\cdot(\|\A^{\! T}(\y-\A\hat{\x})\|_{\infty}\nonumber\\
&&+\|\A^{\! T}(\y-\A\x)\|_{\infty}+\|\A^{\! T}\A\x_{-\max(k)}\|_{\infty})\nonumber\\
&\le&\sqrt{(s^q+1)k}\|\h_{\max(k)}+\h_1\|_2\nonumber\\
&&\cdot(\eta+\epsilon+\|\A^{\! T}\A\x_{-\max(k)}\|_2)\nonumber\\
&\le&\sqrt{(s^q+1)k}\|\h_{\max(k)}+\h_1\|_2 \nonumber\\
&&\cdot(\eta+\epsilon+\sigma^2(\A)\|\x_{-\max(k)}\|_2)\nonumber.
\end{eqnarray}
In this case, the assumption $\|\y-\A\x_T\|_2\in\cB$ holds if $\eta\ge\epsilon+\sigma^2(\A)\|\x_{-\max(k)}\|_2$. Therefore, in \eqref{eq:proof2}, let $\rho=\sqrt{1+\delta}(\epsilon+\eta+\sigma^2(\A)\|\x_{-\max(k)}\|_2)$, then  we have \eqref{eq:th3} from \eqref{eq:proof3}. This finishes the proof of Theorem \ref{thm:3}.
 \end{enumerate}
\end{proof}

\section{\label{sec:conclusion}Conclusion}
We improved the RIC bound of $\ell_q$ minimization by generalizing the result in Cai and Zhang \cite{cai2013sparse}. Under the more general RIC bound, $\ell_q$ minimization can recover sparse signals exactly and approximately sparse signals stably. Although it is a step forward for the RIC study of $\ell_q$ minimization, whether the proposed bound is sharp or not needs further research.
\bibliographystyle{IEEETran}
\bibliography{Bib}

\end{document}